\documentclass[12pt]{article}

\usepackage{adjustbox,multirow,url,bbm,amsmath,amssymb,amsfonts,
amssymb,amsmath,graphicx,setspace
}
\usepackage[a4paper, lmargin=0.1\paperwidth, rmargin=0.1\paperwidth, tmargin=0.1\paperheight, bmargin=0.1\paperheight]{geometry}

\begin{document}

\title{A general approach to detect gene (G)-environment (E) additive interaction leveraging G-E independence in case-control studies}

\author{Eric J. Tchetgen Tchetgen$^1$, Xu Shi$^2$, Tamar Sofer$^3$, Benedict H.W. Wong$^2$ \\ 
	\em $^1$The Wharton School of Business, University of Pennsylvania\\
	\em $^2$Department of Biostatistics, Harvard T.H. Chan School of Public Health\\
	\em $^3$Brigham and Women's Hospital; Harvard Medical School
}
\date{}\maketitle

	\begin{abstract}
It is increasingly of interest in statistical genetics to test for the presence of a mechanistic interaction between genetic (G) and environmental (E) risk factors by testing for the presence of an additive G$\times$E interaction. In case-control studies involving a rare disease, a statistical test of no additive interaction typically entails a test of no relative excess risk due to interaction (RERI). It is also well known that a test of multiplicative interaction exploiting G-E independence can be dramatically more powerful than standard logistic regression for case-control data. Likewise, it has recently been shown that a likelihood ratio test of a null RERI incorporating the G-E independence assumption (RERI-LRT) outperforms the standard RERI approach. In this paper, the authors describe a general, yet relatively straightforward approach to test for G$\times$E additive interaction exploiting G-E independence. The approach which relies on regression models for G and E is particularly attractive because, unlike the RERI-LRT, it allows the regression model for the binary outcome to remain unrestricted. Therefore, the new methodology is completely robust to possible mis-specification in the outcome regression. This is particularly important for settings not easily handled by RERI-LRT, such as when E is a count or a continuous exposure with multiple components, or when there are several auxiliary covariates in the regression model. While the proposed approach avoids fitting an outcome regression, it nonetheless still allows for straightforward covariate adjustment. The methods are illustrated through an extensive simulation study and an ovarian cancer empirical application.
\end{abstract}

\maketitle
\clearpage
\begin{spacing}{1.3}
\section{Introduction}
There is growing interest in the development and application of
statistical methods to detect the presence of an additive gene (G)-environment
(E) interaction because such interaction may be closer to a true mechanistic
interaction than its multiplicative counterpart \cite{rothman1980concepts,greenland1983tests,
cordell2002epistasis,vanderweele2014tutorial,liu2017robust}. For
case-control data involving a rare disease, a statistical test of no additive
G$\times$E interaction is easily performed via a test of a null relative excess risk
due to interaction (RERI) \cite{rothman2008modern}. This approach has gained
popularity in epidemiology primarily because it is easily performed using
relative risk estimates from a standard logistic regression for case-control
data \cite{rothman2008modern}. When G and E are known to be independent in the
target population, it is well known that a test of multiplicative interaction
incorporating the independence assumption can be dramatically more powerful
than standard logistic regression, which does not make use of the assumption
\cite{piegorsch1994non,umbach1997designing,chatterjee2005semiparametric,tchetgen2010semiparametric,tchetgen22robust}. Likewise,
it was recently shown that a likelihood ratio test of the null hypothesis of
no RERI incorporating the G-E independence assumption (hereafter RERI-LRT)
generally outperforms the standard RERI test of no additive interaction \cite{han2012likelihood}. Notably, RERI-based tests of additive interaction rely on
correct specification of a logistic model for the binary outcome, as a
function of G, E and auxiliary covariates. Ideally, a nonparametric
specification of the outcome logistic regression would in principle ensure
validity of RERI-based tests. However, nonparametric estimation may not be
feasible if, as often the case in practice, the environmental exposure or
auxiliary covariates in the model include multiple factors, including count or
continuous variables. Thus, in practice it is customary to specify a
parametric logistic outcome model, therefore producing a parametric test of
the null hypothesis of no RERI. Unfortunately, as we argue in Section~\ref{fail}, parametric
RERI test statistics based on a standard specification of a logistic outcome
regression, will a priori rule out the null hypothesis of no additive
interaction in most practical situations where at least one exposure is either
a count or continuous, therefore leading to inflation of its type 1 error
rate. The presence of covariates has previously been noted as also potentially
problematic for RERI by \cite{skrondal2003interaction}. He has argued, quite convincingly
that given a conceptualization of interaction as departure from additive
risks, making direct inferences regarding the fundamental additive interaction
parameter would be preferred to the common indirect strategy based on RERI, in
order to avoid potential bias due to model misspecification of the outcome
regression. We therefore re-iterate Skrondal's warning against the
indiscriminate use of parametric RERI-based tests of interaction in settings
where saturated or nonparametric specification of the outcome model is not
practical.

In this paper, the authors present a general, yet fairly straightforward
approach to directly test for the presence of additive G$\times$E interaction in
case-control studies without requiring a regression model of disease risk. The
proposed approach which is easily made to exploit the G-E independence
assumption leading to dramatic increase in power, relies on separate
regression models for G and E given covariates. By avoiding specification of
an outcome model, the approach circumvents aforementioned difficulties with
RERI-based tests and is completely robust to mis-specification of the outcome
regression. However, as noted above, correct specification of models for $G$
and $E$ is instead needed for the new approach to be valid.  Nonetheless,
unlike RERI-based tests, standard parametric models can be used for the latter
in most practical situations, even if $E$ is continuous without a priori ruling
out the null hypothesis of no additive interaction. The methods are
illustrated through an extensive simulation study in the simple setting of
binary G and E with no additional covariates so that RERI and the new approach
equally apply and can be directly compared in terms of power. Additional
simulations are performed to illustrate the poor behavior of parametric
RERI-based tests in more practical scenarios.  Next, we demonstrate the new
approach using data from an ovarian cancer study to detect an additive interaction
between the BRCA1/2 genetic variant (G), and the woman's parity and number of
years of oral contraceptive use (E). Because both environmental exposures are
counts, the RERI-LRT cannot easily be implemented without possibly recoding
the original environmental exposures as dichotomous or as categorical with few
levels. Covariate adjustment needed in the study also presents additional difficulty for
RERI-LRT and for these reasons the approach is forgone in both applications in
favor of the new methodology.

\section{Alternative characterization of test of additive interaction}

\noindent Suppose one has observed case-control data on $n$ unrelated
individuals, let $D$ denote the rare disease outcome defining case-control
status and $\left(  A_{1},A_{2}\right)  $ denote two exposures in view. For
instance, in a statistical genetic application, $A_{1}$ may denote the genetic
variant G and $A_{2}$ an environmental exposure E, however, we will use the
more generic notation $\left(  A_{1},A_{2}\right)  $ to allow for more general
contexts considered in Section~\ref{unified}, say where either or both exposures may be count or continuous. Let $\mu(a_{1},a_{2})=\Pr(D=1|A_{1}=a_{1},A_{2}%
=a_{2})$ denote the disease risk of individuals in the target population with
exposure values ($a_{1}$, $a_{2})$. 

\subsection{Binary exposures}
In the case of binary genetic variant and environmental exposure, we have the following saturated model
\[
\mu(A_{1},A_{2})=\beta_0+\beta_1A_1+\beta_2A_2+\beta_3A_1A_2.
\]
Therefore, an
additive interaction between $A_{1}$ and $A_{2}$ is said to be present if
\[
\beta_{3}=\mu(1,1)-\mu(1,0)-\mu(0,1)+\mu(0,0)\neq0,
\]
or equivalently if $RERI\neq0,$ where
\begin{align*}
RERI &  =\left\{  \mu(1,1)-\mu(1,0)-\mu(0,1)\right\}  /\mu(0,0)+1\\
&  =\beta_{3}/\mu(0,0).
\end{align*}
An empirical version of RERI is obtained under case-control sampling by
estimating the required risk ratios $\mu(a_{1},a_{2})/\mu(0,0),$ via a
saturated logistic regression under the rare disease assumption. Then,
standardizing the empirical estimate $\widehat{RERI}$ by a consistent estimate
of its standard error $\sqrt{\widehat{\sigma}_{RERI}^{2}}$ gives the RERI test
statistic $T_{RERI}=\widehat{RERI}/\sqrt{\widehat{\sigma}_{RERI}^{2}}$. It can
then be showed using standard asymptotic arguments that under the null
hypothesis we wish to test, of no additive interaction $H_{0}:\beta
_{3}=RERI=0,$ $T_{RERI}$ is approximately standard normal in large samples.

The following result gives an alternative characterization of the null
hypothesis of no additive interaction which motivates the new approach. To
state the result, let $\pi_{1}\left(  a_{2}\right)  =\Pr(A_{1}=1|A_{2}=a_{2})$
denote the prevalence of the first exposure $A_{1}$ among individuals with the
second exposure $A_{2}=a_{2}$ in the underlying population, and likewise
define $\pi_{2}(a_{1})=\Pr(A_{2}=1|A_{1}=a_{1})$. Also let $\alpha$ denote the
log odds ratio association relating $A_{1}$ and $A_{2}$ in the target
population, thus
\[
\exp\alpha=\frac{\pi_{1}\left(  1\right)  (1-\pi_{1}\left(  0\right)  )}%
{\pi_{1}\left(  0\right)  (1-\pi_{1}\left(  1\right)  )}\
\]
such that $\alpha=0$ encodes the independence assumption between $A_{1}$ and
$A_{2}.$

\noindent\textbf{Result 1. }\textit{We have that the null hypothesis of no
additive interaction }$H_{0}$ \textit{holds if and only if}
\[
RERI=0\Leftrightarrow\mathbb{E}\left\{  U|D=1\right\}  =0
\]
\textit{where }%
\[
U=e^{-\alpha A_{1}A_{2}}\left(  A_{1}-\pi_{1}(0)\right)  \left(  A_{2}-\pi
_{2}(0)\right)  D.
\]

We should note that Result 1 does not rely on the rare disease assumption and
holds irrespective of the population disease prevalence. The result is a
special case of a more general lemma given later in the text allowing for
arbitrary exposures and for covariate adjustment. According to the result, the
null hypothesis of no additive interaction holds if and only if RERI is equal
to zero, or equivalently if and only if the random variable $U$ has mean zero among cases ($D=1$).
Intuition about the result is gained by assuming G-E independence, i.e.
$\alpha=0,$ such that $\pi_{j}(a)=\pi_{j}$. Then, upon noting that the
conditional density of $(A_{1},A_{2})$ given $D=1$ is proportional to
\[
\mu(A_{1},A_{2})f_{1}(A_{1})f_{2}(A_{2})=(\beta_{0}+\beta_{1}A_{1}+\beta
_{2}A_{2}+\beta_{3}A_{1}A_{2})f_{1}(A_{1})f_{2}(A_{2})
\]
where $f_{j}(1)=\pi_{j}$, one observes that $\mathbb{E}\left\{  U|D=1\right\}
$ is proportional to
\begin{align*}
&  \sum_{a_{1},a_{2}}\left(  a_{1}-\pi_{1}\right)  \left(  a_{2}-\pi
_{2}\right)  (\beta_{0}+\beta_{1}a_{1}+\beta_{2}a_{2}+\beta_{3}a_{1}%
a_{2})f_{1}(a_{1})f_{2}(a_{2})\\
&  =\beta_{0}\underset{=0}{\underbrace{\sum_{a_{1},a_{2}}\left(  a_{1}-\pi
_{1}\right)  \left(  a_{2}-\pi_{2}\right)  f_{1}(a_{1})f_{2}(a_{2})}}\\
&  +\beta_{1}\underset{=0}{\underbrace{\sum_{a_{1},a_{2}}\left(  a_{1}-\pi
_{1}\right)  \left(  a_{2}-\pi_{2}\right)  a_{1}f_{1}(a_{1})f_{2}(a_{2})}%
}+\beta_{2}\underset{=0}{\underbrace{\sum_{a_{1},a_{2}}\left(  a_{1}-\pi
_{1}\right)  \left(  a_{2}-\pi_{2}\right)  a_{2}f_{1}(a_{1})f_{2}(a_{2})}}\\
&  +\beta_{3}\sum_{a_{1},a_{2}}\left(  a_{1}-\pi_{1}\right)  \left(  a_{2}%
-\pi_{2}\right)  a_{1}a_{2}f_{1}(a_{1})f_{2}(a_{2})\\
&  =\beta_{3}\pi_{1}\left(  1-\pi_{1}\right)  \pi_{2}\left(
1-\pi_{2}\right),
\end{align*}
confirming that $\mathbb{E}\left\{  U|D=1\right\}  =0$ if and only if the
additive interaction $\beta_{3}=0.$ Result 1 further shows that a similar
result holds when the exposures are dependent upon applying a weight to
individuals with both exposures, equal to the inverse of the odds ratio
association of the two exposures. Intuitively, weighting makes the exposures
independent, thus essentially recovering the independent exposure setting in
the weighted sample. Since $U$ only uses exposure data among cases (with
$D=1)$, the result suggests that one may be able to test for additive
interaction by considering whether the distribution of the exposures in view
satisfies the above condition using data for cases only. Unfortunately, $U$ is
not directly observed and therefore cannot directly be used for inference, as
it depends on the unknown population parameters $\pi_{j}(0),$ $j=1,2.$
Nonetheless, progress can be made under the rare disease assumption, since one
may use the controls (with $D=0)$ for approximate inference, upon observing
that $\pi_{j}(0)\approx p_{j}(0)$ where $p_{1}(a_{2})=\Pr(A_{1}=1|A_{2}%
=a_{2},D=0)$ and $p_{2}(a_{1})=\Pr(A_{2}=1|A_{1}=a_{1},D=0).$ 
Specifically, let $\omega=\log[p_{1}(1)(1-p%
_{1}(0))/p_{1}(0)(1-p_{1}(1))]$, then $\omega \approx \alpha$ under the rare disease assumption.
Therefore, one
may estimate $\sum_{i}U_{i}$ with $\sum_{i}\widehat{U}_{i}$ where%
\[
\widehat{U}_{i}=\exp\left(  -A_{1,i}A_{2,i}\widehat{\omega}\right)  \left(
A_{1,i}-\widehat{p}_{1}(0)\right)  \left(  A_{2,i}-\widehat{p}_{2}(0)\right)
D_{i},
\]
with $\widehat{p}_{1}(a)=$ $\sum_{i}A_{1,i}I(A_{2,i}=a,D=0)/$ $\sum
_{i}I(A_{2,i}=a,D=0)$ the sample version of $p_{1}(a)$, $\widehat{p}_{2}(a)$
similarly defined, and exp($\widehat{\omega})=\widehat{p}_{1}(1)(1-\widehat{p}%
_{1}(0))/\widehat{p}_{1}(0)(1-\widehat{p}_{1}(1))$ the sample odds ratio
relating $A_{1}$ and $A_{2}$ in the controls$.$ In the Appendix, we show how
to derive $\sigma_{\widehat{U}}^{2}=Var(\sum_{i}\widehat{U}_{i}/n)$ (see
equation $\left(  1\right)  $ of the Appendix)$.$ Suppose that unbeknownst to
the analyst, $A_{1}$ and $A_{2}$ are independent in the population and
therefore $\widehat{\omega}$ converges to $0$ in probability. We evaluate
$\sigma_{\widehat{U}}^{2}$ at this particular submodel and show that
$\sigma_{\widehat{U}}^{2}$ can be decomposed as $\widehat{\sigma}%
_{\widehat{U}}^{2}$ $=\widehat{V}_{1}+\widehat{V}_{2}+\widehat{V}_{3},$ where
$\widehat{V}_{j}$ is an estimate of $V_{j}$, $j=1,2,3,$ described in the
Appendix. Considering in turn each contribution to the variance, we note that
the first term $\widehat{V}_{1}$ captures the variance of $\sum_{i}U_{i}/n$ if
$\left(  \omega,p_{1}(0),p_{2}(0)\right)  $ were known; the second term
$\widehat{V}_{2}$ reflects the uncertainty due to estimation of $(p_{1}%
(0),p_{2}(0));$ while $\widehat{V}_{3}$ reflects the uncertainty associated
with estimation of the odds ratio parameter $\omega.$ In the next section, we further
consider how the explicitly leveraging G-E independence assumption alters each of these
contributions to reveal how the assumption can improve power to detect the
presence of an additive interaction. 

Here we note that, under $H_{0}$ the
standardized test statistic $T=\sum_{i}\widehat{U}_{i}/\left(  n\sqrt
{\widehat{\sigma}_{\widehat{U}}^{2}}\right)  $ is approximately standard
normal in large samples. Under the two-sided alternative hypothesis $\beta
_{3}\neq0$, one can further show that in large samples, $T$ has approximate
variance one, and is approximately centered at the non-centrality parameter
$\kappa\times\beta_{3},$ where:
\[
\kappa=p_{1}(0)\left(  1-p_{1}(0)\right)  p_{2}(0)\left(  1-p_{2}(0)\right)
\lambda/\sigma_{\widehat{U}}^{2},
\]
$\lambda$ is the sampling fraction of cases (i.e. $\lambda=$ proportion of
cases in case-control sample/proportion of cases in population). Thus, $T$ has
asymptotic power one since $1/\sigma_{\widehat{U}}^{2}$ and therefore $\kappa$
tends to infinity with sample size; confirming that similar to $T_{RERI},$ $T$
is a consistent test statistic of $H_{0}$. 

Interestingly, the above derivation also implies that the statistic 
$\sum_{i}\widehat{U}_{i}/\{
\widehat{p}_{1}\left(  1-\widehat{p}_{1}(0)\right)  \widehat{p}_{2}(0)(
1-\widehat{p}_{2}(0))  \sum_{i}D_{i}\}$ gives a consistent
estimate of $\beta_{3}/\Pr\{D=1\}$ the interaction parameter of interest
scaled by the inverse of the population disease prevalence. Thus, one could in
principle recover a consistent estimate of $\beta_{3}$ if either the
underlying population disease prevalence or the sampling fraction of cases
were known. 

We note that neither $T$ nor $T_{RERI}$ makes explicit use of the G-E
independence assumption and therefore both may be inefficient when the
assumption holds. In the following section, we modify $T$ to explicitly
encode the independence assumption thus obtaining a more powerful test statistic.

\subsection{Test incorporating independence assumption}

\noindent Suppose that $A_{1}$ and $A_{2}$ are known to be independent in the
population. Naturally, one may wish to exploit such prior information in
testing for G-E interaction. This can be accomplished by adapting the
methodology developed in the previous section upon noting that the
independence assumption implies $\alpha=0,$ which, under the rare disease
assumption, also implies that $\omega\approx0.$ This leads us to modify
$\widehat{U}_{i}$. Define $\widetilde{U}_{i}$ similarly to $\widehat{U}_{i}$
with $\widehat{\omega}=0,$ i.e. 
\[
\widetilde{U}_{i}=\left(  A_{1,i}%
-\widehat{p}_{1}(0)\right)  \left(  A_{2,i}-\widehat{p}_{2}(0)\right)  D_{i}.\]
In the appendix, we show that $\sigma_{\widetilde{U}}^{2}=Var(\sum
_{i}\widetilde{U}_{i}/n)$ can be estimated by $\widehat{\sigma}_{\widetilde{U}%
}^{2}$ $=\widehat{V}_{1}+\widehat{V}_{2}.$ Consequently $\widehat{\sigma
}_{\widetilde{U}}^{2}<\widehat{\sigma}_{\widehat{U}}^{2}$, reflecting the
efficiency gain due to the independence assumption, i.e. $\widehat{V}_{3}$ is
exactly zero since there is no uncertainty associated with $\widehat{\omega
}=0$. One can verify that the non-centrality parameter $\beta_{3}\times
\kappa_{1}$ of $T_{1}=\sum_{i}\widetilde{U}_{i}/n\sqrt{\widehat{\sigma
}_{\widetilde{U}}^{2}}$ becomes $\kappa_{1}=\frac{\sigma_{\widehat{U}}}%
{\sigma_{\widetilde{U}}}\kappa>\kappa$, confirming that $T_{1}$ is guaranteed
to be more powerful than $T$.

\subsection{ Adjusting for covariates}

\noindent In observational studies, it is usually desirable to adjust for
potential confounding of the joint effects of $A_{1}$ and $A_{2},$ and such
covariate adjustment may also be required to enforce the G-E independence
assumption$.$ Let $X$ denote such a vector of covariates and suppose that the
exposures are independent conditional on $X.$ Define $p_{1}(x)=\Pr
(A_{1}=1|X=x,D=0)$ and $p_{2}(x)=\Pr(A_{2}=1|X=x,D=0).$ Likewise, let
$\widehat{p}_{1}(x)$ and $\widehat{p}_{2}(x)$ correspond to estimates,
obtained using standard parametric models, e.g. using logistic regressions of
the form logit$\widehat{p}_{j}(x)=$logit$p_{j}(x;\widehat{\theta}%
_{j})=(1,x^{\prime})\widehat{\theta}_{j},j=1,2,$ computed by maximum
likelihood. The test statistic $T_{2}=\sum_{i}\overline{U}_{i}/\sqrt
{\widehat{\sigma}_{\overline{U}}^{2}}$ has under the null hypothesis of no
additive interaction, an approximate standard normal distribution, with
$\overline{U}_{i}$ defined as
\[
\overline{U}_{i}=\ \left(  A_{1,i}-\widehat{p}_{1}(X)\right)  \left(
A_{2,i}-\widehat{p}_{2}(X)\right)  D_{i},
\]
where $\widehat{\sigma}_{\overline{U}}^{2}\ $is obtained using equation
$\left(  1\right)  $ of the Appendix.

\section{ More general exposures }

\noindent Next, suppose that the environmental exposure $A_{2}$ were
continuous, for example if $D$ were diabetes status, $A_{2}$ could be body
mass index (BMI) typically coded on a continuous scale. Note that the null
hypothesis of no additive interaction can be restated as followed to
acknowledge the continuous exposure:%
\[
H_{0}:\mu(1,a_{2},x)-\mu(1,0,x)-\mu(0,a_{2},x)+\mu(0,0,x)=0\text{ for all
values of }a_{2}\text{ and }x,
\]
where $\mu(a_{1},a_{2},x)=\Pr(D=1|a_{1},a_{2},x)$. \ To construct an
appropriate test statistic of $H_{0}$, suppose that $\mathbb{E}\left(
A_{2}|X=x,D=0\right)  $ is estimated with the linear model $\widehat{m}%
_{2}(x)=$ $m_{2}(x,\widehat{\theta}_{2})=(1,x^{\prime})\widehat{\theta}_{2}$
via ordinary least squares using controls only. Assuming G-E conditional
independence given $X$, it is straightforward to modify the proposed test
statistic to account for the continuous exposure, by simply replacing
$\widehat{p}_{2}(x)$ with $\widehat{m}_{2}(x).$ Thus, we let
\[
\overline{U}_{i}^{c}=\ \left(  A_{1,i}-\widehat{p}_{1}(X_{i})\right)  \left(
A_{2,i}-\widehat{m}_{2}(X_{i})\right)  D_{i},
\]
and $\widehat{\sigma}_{\overline{U}^{c}}^{2}$ denotes an estimate of the
variance of $\sum_{i}\overline{U}_{i}^{c}/n$ obtained using equation $\left(
1\right)  $ of the Appendix. Then, the test statistic $T_{3}=\sum_{i}%
\overline{U}_{i}^{c}/n\sqrt{\widehat{\sigma}_{\overline{U}^{c}}^{2}}$ is
approximately standard normal under $H_{0}$. 

A similar test statistic could be
defined if $A_{2}$ were a count, upon estimating its mean with the log-linear
model $\log n_{2}(x,\widehat{\theta}_{2})=(1,x^{\prime})\widehat{\theta}_{2}$
computed by maximum likelihood under say a Poisson model for $A_{2}$. Then,
one could simply replace $\widehat{m}_{2}$ with $\widehat{n}_{2}$ in defining
the test statistic, and one could likewise modify the estimated variance of
the test statistic using $\left(  1\right)  $ of the Appendix.

In order to simplify the presentation, thus far we have taken $A_{1}$ to be a
binary genetic variant. Suppose now that $A_{1}$ were more generally
categorical having $K$ possible levels $\{0,a_{1,1},...,a_{1,K-1}\}$ with $0$
a reference value. For instance, if $A_{1}$ were to encode the number of minor
alleles measured at a single nucleotide polymorphism (SNP) locus, then $K=3,$
and $a_{1,k}=k,$ $k=1,2.$ Further assuming say that $A_{2}$ were continuous
and independent of $A_{1}$ given $X$, we could then simply define
\[
\overline{U}_{i}^{m}=\ \sum_{k=1}^{K-1}\left(  I(A_{1,i}=a_{1,k}%
)-\widehat{p}_{1,k}(X_{i})\right)  \left(  A_{2,i}-\widehat{m}_{2}%
(X_{i})\right)  D_{i},
\]
where $\widehat{p}_{1,k}(x)$ is a maximum likelihood estimate of $\Pr
(A_{1}=a_{k}|x)$ computed using standard polytomous logistic regression. Let
$\widehat{\sigma}_{\overline{U}^{m}}^{2}$ denote an estimate of the large
sample variance of $\sum_{i}\overline{U}_{i}^{m}/n$ based on $\left(
1\right)  $ of the Appendix. Then in large samples, the resulting test
statistic $T_{4}=\sum_{i}\overline{U}_{i}^{m}/n\sqrt{\widehat{\sigma
}_{\overline{U}^{m}}^{2}}$ is approximate standard normal under the null
hypothesis of no additive interaction which may be restated to account for the
polytomous and continuous exposures:%
\[
H_{0}:\mu(a_{1,k},a_{2},x)-\mu(a_{1,k},0,x)-\mu(0,a_{2},x)+\mu(0,0,x)=0\text{
for all }k,\text{ and all values of }a_{2}\text{ and }x.
\]

\subsection{ Failure of RERI-based approaches with continuous exposure\label{fail}}

\noindent We now describe in some detail, the aforementioned failure of
RERI-based approaches that use standard logistic regression when at least one
exposure is non-discrete and auxiliary covariates are present. In this vein,
suppose that $A_{1}$ is continuous, while $A_{2}$ may be binary. In practice,
to evaluate RERI in this context, one typically proceeds by estimating a
standard logistic regression for $\Pr\left\{  D=1|a_{1},a_{2},x\right\}  $
using a simple parametric formulation of the model, such as:
\begin{equation}
\text{logit}\Pr\left\{  D=1|a_{1},a_{2},x;\alpha_{0},\alpha_{1},\alpha
_{2},\alpha_{3},\alpha_{4}\right\}  =\alpha_{0}+\alpha_{1}a_{2}+\alpha
_{2}a_{1}+\alpha_{3}a_{1}a_{2}+\alpha_{4}^{\prime}x,\label{standard model}%
\end{equation}
where logit$(p)=$log$\left\{  p/(1-p)\right\}  $ and the parameters $\left(
\alpha_{0},\alpha_{1},\alpha_{2},\alpha_{3},\alpha_{4}\right)  $ are variation
independent \cite{knol2007estimating}. Below, we argue that such a standard logistic
regression will generally be incompatible with the null hypothesis of no
additive interaction if both exposures $\left(  a_{1},a_{2}\right)  $ have a
non-null association with the outcome. Specifically, suppose that the main
effects of $A_{1}$ and $A_{2}$ and $X$ are correctly specified in the logistic
model, i.e.
\begin{align*}
\text{logit}\Pr\left\{  D=1|a_{1}=0,a_{2},x\right\}    & =\alpha_{0}%
+\alpha_{1}a_{2}+\alpha_{4}^{\prime}x\\
\text{logit}\Pr\left\{  D=1|a_{1},a_{2}=0,x\right\}    & =\alpha_{0}%
+\alpha_{2}a_{1}+\alpha_{4}^{\prime}x
\end{align*}
with $\alpha_{1}\neq0$ and $\alpha_{2}\neq0.$ Then there will generally be no
parameter value of $\left(  \alpha_{0},\alpha_{1},\alpha_{2},\alpha_{3}%
,\alpha_{4}\right)  $ that encodes the null hypothesis of no additive
interaction, consequently any RERI-type test, based on model $\left(
\ref{standard model}\right)  $ will generally have inflated type 1 error rate
for testing the null of no additive interaction. To further understand the
failure of RERI in this context, note that under the null hypothesis of no
additive interaction
\[
\Pr\{  D\!=\!1|a_{1},a_{2},x\}  \!=\!\Pr\{  D\!=\!1|a_{1}\!=\!0,a_{2}%
,x\}  +\Pr\{  D\!=\!1|a_{1},a_{2}\!=\!0,x\}  -\Pr\{
D\!=\!1|a_{1}\!=\!0,a_{2}\!=\!0,x\}
\]
for all possible values of $a_{1}$ and $a_{2}$. Then under the null, the
interaction function on the log odds ratio scale is given by
\begin{align*}
& \theta\left(  a_{1},a_{2},x;\alpha_{0},\alpha_{1},\alpha_{2},\alpha
_{4}\right)  \\
& =\text{logit}\Pr\left\{  D=1|a_{1},a_{2},x;\alpha_{0},\alpha_{1},\alpha
_{2},\alpha_{4}\right\}  -\text{logit}\Pr\left\{  D=1|a_{1}=0,a_{2}%
,x;\alpha_{0},\alpha_{1},\alpha_{4}\right\}  \\
& -\text{logit}\Pr\left\{  D=1|a_{1},a_{2}=0,x;\alpha_{0},\alpha_{2}%
,\alpha_{4}\right\}  +\text{logit}\Pr\left\{  D=1|a_{1}=0,a_{2}=0,x;\alpha
_{0},\alpha_{4}\right\}  \\
& =\text{logit}\{\Pr\left\{  D=1|a_{1}=0,a_{2},x;\alpha_{0},\alpha_{1}%
,\alpha_{4}\right\}  +\Pr\left\{  D=1|a_{1},a_{2}=0,x;\alpha_{0},\alpha
_{2},\alpha_{4}\right\}  \\
& -\Pr\left\{  D=1|a_{1}=0,a_{2}=0,x;\alpha_{0},\alpha_{4}\right\}  \}\\
& -\text{logit}\Pr\left\{  D=1|a_{1}=0,a_{2},x;\alpha_{0},\alpha_{1}%
,\alpha_{4}\right\}  -\text{logit}\Pr\left\{  D=1|a_{1},a_{2}=0,x;\alpha
_{0},\alpha_{2},\alpha_{4}\right\}  \\
& +\text{logit}\Pr\left\{  D=1|a_{1}=0,a_{2}=0,x;\alpha_{0},\alpha
_{4}\right\},
\end{align*}
in which case, correct specification of a logistic model for $\Pr\left\{
D=1|a_{1},a_{2},x\right\}  $ under a null additive interaction is of the form
\begin{equation}
\text{logit}\Pr\left\{  D=1|a_{1},a_{2},x;\alpha_{0},\alpha_{1},\alpha
_{2},\alpha_{4}\right\}  =\alpha_{0}+\alpha_{1}a_{2}+\alpha_{2}a_{1}%
+\theta\left(  a_{1},a_{2},x;\alpha_{0},\alpha_{1},\alpha_{2},\alpha
_{4}\right)  +\alpha_{4}^{\prime}x.\label{null model}%
\end{equation}
Because of the nonlinear dependence of $\theta$ on $a_{1}$ and $x,$ it is
clear that model $\left(  \ref{null model}\right)  $ cannot be nested in the
standard logistic model $\left(  \ref{standard model}\right)  $, and therefore
the latter cannot be used to obtain a valid test of the null hypothesis of no
additive interaction. 

In order to implement an LRT of additive interaction using the RERI approach,
an analyst would need to carefully specify a model for the odds ratio
interaction, so that model $\left(  \ref{null model}\right)  $ is recovered
under the null of no additive interaction. Such a parametrization of the
outcome regression will characteristically be nonstandard in the sense that
the interaction of the resulting logistic model would need to be explicitly
defined as a function of models for both exposure main effects, and the effect
of covariates. Such a parametrization of a logistic model would seldom
naturally arise in practice purely on scientific basis. Furthermore, one would
generally be unable to easily obtain parameter estimates for such a model
using off-the-shelf statistical software for standard logistic regression,
which completely undermines the often quoted practical advantage of the RERI
approach. 

\section{ A unified class of test statistics\label{unified}}

\noindent We now provide a unified class of test statistics for the null
hypothesis of no additive interaction which subsumes as special case, each of
the settings considered in previous sections, but which also allows for the
conditional independence assumption of the two exposures to be relaxed.

 To do so, we proceed as in \cite{tchetgen2009doubly} and use
the following representation of the joint density of $\left(  A_{1}%
,A_{2}\right)  $ given $X:$%
\begin{equation}
f(A_{1},A_{2}|X)=\frac{f(A_{1}|A_{2}=0,X)f(A_{2}|A_{1}=0,X)OR(A_{1},A_{2}%
;X)}{\int\int f(a_{1}|A_{2}=0,X)f(a_{2}|A_{1}=0,X)OR(a_{1},a_{2};X)d\nu
(a_{1},a_{2})}, \label{joint1}%
\end{equation}
where $\nu$ is a dominating measure of the distribution of $\left(
A_{1},A_{2}\right)  ,$ $OR(A_{1},A_{2};X)$ is the generalized odds ratio
function relating $A_{1}$ and $A_{2}$ within levels of $X,$ that is
\[
OR(A_{1},A_{2};X)=\frac{f(A_{1},A_{2}|X)f(A_{1}=0,A_{2}=0|X)}{f(A_{1}%
=0,A_{2}|X)f(A_{1},A_{2}=0|X)}%
\]
and $\left\{  f(A_{1}|A_{2}=0,X),f(A_{2}|A_{1}=0,X)\right\}  $ are baseline
densities in the target population. Note that the generalized odds ratio
function reduces in the simple case of binary exposures, to the standard odds
ratio effect measure, but remains well defined as a measure of association for
exposures of a more general nature, whether categorical, count or continuous
variables, i.e.\ $OR(A_{1},A_{2};X)=1$ if and only if $A_{1}$ and $A_{2}$ are
independent within levels of $X$. The null hypothesis of no additive
interaction can more generally be stated as:%
\[
H_{0}:\mu(a_{1},a_{2},x)-\mu(a_{1},0,x)-\mu(0,a_{2},x)+\mu(0,0,x)=0\text{ for
all values of }a_{1},a_{2}\text{ and }x.
\]
For any function $g(A_{1},A_{2},X)$ of $(A_{1},A_{2},X),$ let%

\begin{align*}
w(A_{1},A_{2},X,D;g)\noindent &  =W(g)\\
&  =OR(A_{1},A_{2};X)^{-1}\{g(A_{1},A_{2},X)-\int g(A_{1},a_{2},X)f(a_{2}%
|A_{1}=0,X)d\mu(a_{2})\\
&  -\int g\left(  a_{1},A_{2},X\right)  f\left(  a_{1}|A_{2}=0,X\right)
d\mu(a_{1})\\
&  +\int g(a_{1},a_{2},X)f\left(  a_{2}|A_{1}=0,X\right)  f\left(  a_{1}%
|A_{2}=0,X\right)  d\mu(a_{1},a_{2})\}D.
\end{align*}

\noindent\textbf{Lemma 1. }\textit{The null hypothesis }$\mathit{H}_{0}$
\textit{holds if and only if }%
\[
\mathbb{E}\left\{  W(g)|D=1,x\right\}  =0\text{ \textit{for all values of}
}\mathit{x}\text{ \textit{and all functions} }g.
\]

Result 1 is easily recovered as a corollary of Lemma 1. According to
Lemma 1, an empirical version of $W(g)$ with user-specified function $g$ may be
used to test $H_{0}.$ One must estimate the unknown odds ratio function and
baseline densities, in order to obtain an estimate of the joint density of
$(A_{1},A_{2})$ given $X.$ Under the rare disease assumption, estimation of
the joint density can proceed by standard maximum likelihood in the controls
only, using the parametrization given in equation $\left(  \ref{joint1}%
\right)  $, upon positing parametric models for the odds ratio function and
baseline densities$.$ To ground ideas, suppose one posits parametric models
$OR(A_{1},A_{2};X;\omega),f(A_{1}|A_{2}=0,X;\alpha_{1})$ and $f(A_{2}%
|A_{1}=0,X;\alpha_{2}),$ e.g. a single parameter model $\log OR(A_{1}%
,A_{2};X)=\omega A_{1}A_{2}$ may be used that encodes the assumption that the
odds ratio association between $A_{1}$ and $A_{2}$ given $X$ does not vary
with $X$, i.e. no effect heterogeneity in $X$ of the odds ratio association
between $A_{1}$ and $A_{2}$ in the population. For exposures that are either
binary, continuous or counts, generalized linear models within the exponential
family may be used to model the baseline densities. For example, counts may be
modeled by assuming a Poisson distribution for the corresponding baseline
density. Let $\widehat{\omega},$ $\widehat{f}(A_{1}|A_{2}=0,X)$ and
$\widehat{f}(A_{2}|A_{1}=0,X)$ denote the approximate maximum likelihood
estimate of $\left(  \ref{joint1}\right)  $ using controls only; and let
$\widehat{W}(g)=W(g,\widehat{\theta})$ denote the resulting estimate of
$W(g),$ where $\theta=\left(  \omega,\alpha_{1},\alpha_{2}\right)  $. Our
proposed test statistic is then given by $Z=\sum_{i}\widehat{W}_{i}%
(g)/n\sqrt{\widehat{\sigma}_{W}^{2}},$ where $\widehat{\sigma}_{W}^{2}$ is the
estimate of $Var\left(  \sum_{i}\widehat{W}_{i}(g)/n\right)  $ provided in the Appendix.

It is straightforward to verify that the test statistics considered in
previous section belong to the above unifying class of test statistics. For
instance, the test statistics proposed to handle binary, continuous or count
exposures under the independence assumption are obtained by taking:
\[
g(A_{1},A_{2},X)=\left(  A_{1}-{\mathbb{E}}\left(  A_{1}|X\right)
\right)  \left(  A_{2}-{\mathbb{E}}\left(  A_{2}|X\right)  \right),
\]
where ${\mathbb{E}}\left(  A_{j}|X\right)  $ is the mean of $A_{j}$
evaluated under ${f}(A_{j}|X),$ $j=1,2$. For $A_{1}$ categorical with
$K\,$distinct categories\ and $A_{2}$ binary, continuous or a count, one
likewise obtains the test statistic previously proposed by taking:%
\[
g(A_{1},A_{2},X)=\sum_{k=1}^{K-1}\left(  I(A_{1,i}=a_{1,k}%
)-{\mathbb{E}}\left(  I(A_{1,i}=a_{1,k})|X\right)  \right)  \left(
A_{2}-{\mathbb{E}}\left(  A_{2}|X\right)  \right).
\]
Under the independence assumption, the asymptotic variance of $Var\left(
\sum_{i}\widehat{W}_{i}(g)/n\right)  $ is easily modified to account for the
assumption that $OR(A_{1},A_{2};X)$ is set to $1$ for all persons in the sample.

\noindent

\section{Relaxing the rare disease assumption}

\noindent In case the rare disease assumption does not apply, estimating
exposure regression models in controls only may not be entirely appropriate.
Nonetheless, it may still be possible to test for the presence of an additive
interaction, for instance if as often the case in nested case-control studies,
sampling fractions for cases and controls were known. Then, standard inverse
probability weighting could be used based on known sampling weights to
estimate population models for the exposures using both cases and controls.
Potentially more efficient estimates of models for the exposures could
alternatively be obtained using more recent methodology for regression
analysis of secondary outcomes in case-control studies \cite{tchetgen2013general}.

\section{ A simulation study}

\noindent We study the power and type 1 error of our proposed test in the
standard setting of binary genetic and environmental variables with no other
covariate, so that it is more easily compared to the approach of \cite{han2012likelihood}. In order to evaluate both type 1 error rates and power of various test
statistics, we generated simulated data following the design of \cite{han2012likelihood} which encodes the magnitude of the interaction indirectly by varying
RERI from 0 (to assess type 1 error) to 0.5. The probability of having the
genetic variant was 0.5, and the probability of the binary environmental
variable was 0.2, and these factors were generated to be independent. Let
expit$(z)=\exp(z)/[\exp(z)+1]$ and logit$(p)=\log[p/(1-p)]$. The disease risk model was%
\[
\text{logit}\Pr(D=1|a_{1},a_{2})=\alpha_{0}+\alpha_{1}a_{1}+\alpha_{2}%
a_{2}+\alpha_{3}a_{1}a_{2};
\]
with baseline risk equal to $0.01$ (i.e. $\alpha_{0}=$ logit$(0.01)$), the
gene and environment main effects were varied so that $\left(  \alpha
_{1},\alpha_{2}\right)  \in\left\{  \log(0.7),\log(1.2), \log(2)\right\}  $,
and the multiplicative G-E interaction parameter $\alpha_{3}$ was selected to
yield the desired RERI, according to the formula%
\[
\alpha_{3}=\text{logit}[(\text{RERI}-1)\text{expit}(\alpha_{0})+\text{expit}%
(\alpha_{0}+\alpha_{1})+\text{expit}(\alpha_{0}+\alpha_{2})]-\alpha_{0}%
-\alpha_{1}-\alpha_{2}.%
\]
In each simulation, we generate 4000 cases and 4000 controls. We report
results for 10,000 simulations for each setting corresponding to a particular
combination of $\left(  \alpha_{1},\alpha_{2}\right)  $ and RERI values.

Figure 1 summarizes results in terms of power comparing the
proposed tests with and without using the G-E independence assumption, labeled
`U ind' and `U' respectively. The figure also presents results for the
retrospective profile likelihood ratio test proposed by  \cite{han2012likelihood}  with
and without using the independence assumption respectively, labeled `Han ind'
and `Han' respectively. Finally, the figure also displays results from the
standard RERI test based on prospective logistic regression, which is labeled
`prosp'. Table 1 summarizes the type I error rate of the various methods under ranging parameter values.
 
One observes that the RERI-LRT
test `Han ind' and `U ind' are equally powerful when $\Pr(G=1)=0.5$ across
various possible values for the other parameters, and both tests are
dramatically more powerful when compared to the other tests, while `U' is
slightly less powerful than `Han', which is in turn slightly less powerful
than `prosp'.

In additional simulations, we varied the prevalence of the genetic marker
$\Pr(G=1)$ to have population probabilities 0.2 and 0.05, while the
environmental factor was maintained to have probability 0.2. All tests appear to have correct type 1 error rate as shown in Table~1. Power plots
similar to those appearing in Figure 1 are provided in the supplementary
material for these additional settings.
These additional simulations confirm
that all tests become less powerful as the genetic variant becomes less
common, with `Han ind' being slightly more powerful than `U ind' when
$\Pr(G=1)=0.05$.\ Overall, the simulation study confirms that the proposed
approach performs quite competitively when compared with the efficient
RERI-LRT approach, in settings where both methods are available.

In the following section, we consider a data application of the new approach
in which RERI is no longer readily available and cannot easily be applied
without further making unnecessary assumptions.

\begin{figure}[htp]
\begin{centering}\label{fig}
\makebox[\textwidth][c]{\includegraphics[width=1\textwidth, scale=.4]{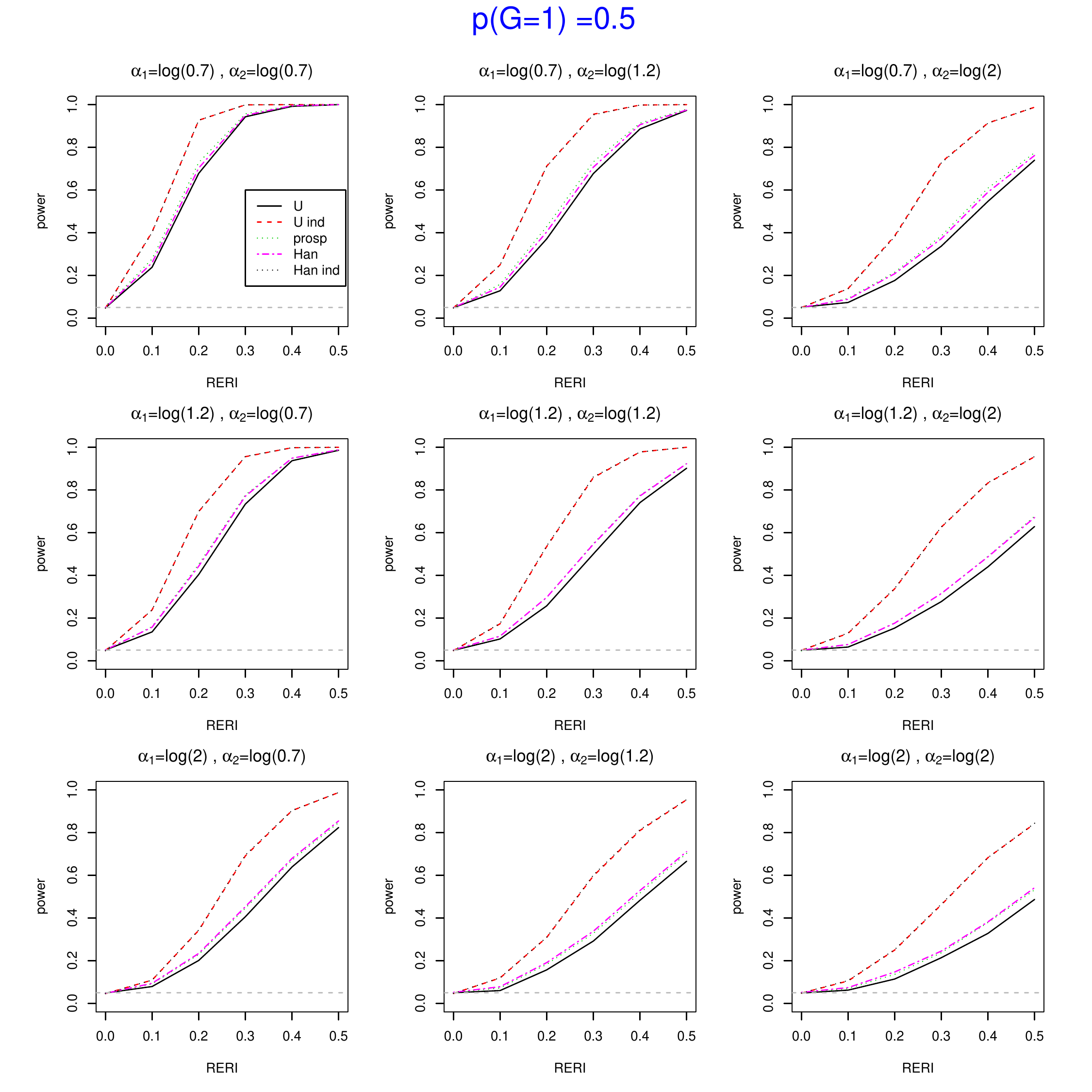}}
\end{centering}
\caption{ Power of the various tests for identifying additive G-E interaction, in the simple settings in which both $a_1$ and $a_2$ are binary, and no covariates are used in the model. The simulated $a_1$ is common $\Pr(a_1=1) =0.5$, $\Pr(a_2=1) =0.2$, and the disease model $\Pr(D=1|a_1,a_2)$ has the fixed baseline risk $\alpha_0 =\mbox{logit}(0.01)$, the main effects of the $a_1, a_2$ are varied as $\alpha_1,\alpha_2$, and the RERI is varied from 0 to 0.5. The compared estimators are the proposed `U' and `U ind' (without and with assuming G-E independence), `prosp' is the standard test under prospective logistic regression and `Han' and `Han ind' is the retrospective profile likelihood test proposed in  \cite{han2012likelihood}  without and with assuming G-E independence. 
}
\end{figure}

\begin{table}[pth]
\label{tab:size} \centering
\begin{tabular}{lrrrrrrrrr}\hline

\hline\hline
\multicolumn{1}{l}{$\alpha_1:$}&\multicolumn{3}{c}{$ \log(0.7)$}&\multicolumn{3}{|c|}{ $\log(1.2)$}&\multicolumn{3}{c}{$\log(2)$}\tabularnewline 
\hline
%
\multicolumn{1}{l}{$\alpha_2:$}&\multicolumn{1}{c}{$\log(0.7)$}&\multicolumn{1}{c}{$\log(1.2)$}&\multicolumn{1}{c}{$ \log(2)$}&\multicolumn{1}{c}{ $ \alpha_2 =\log(0.7)$}&\multicolumn{1}{c}{$  \log(1.2)$}&\multicolumn{1}{c}{$ \log(2)$}&\multicolumn{1}{c}{$ \log(0.7)$}&\multicolumn{1}{c}{$ \log(1.2)$}&\multicolumn{1}{c}{$ \log(2)$}\tabularnewline
\hline
\multicolumn{10}{c}{$\Pr(G=1) = 0.5$} \\ \hline
 U&$0.048$&$0.049$&$0.051$&$0.050$&$0.048$&$0.049$&$0.049$&$0.050$&$0.048$\tabularnewline
 U ind&$0.047$&$0.048$&$0.049$&$0.047$&$0.047$&$0.049$&$0.046$&$0.046$&$0.048$\tabularnewline
 prosp&$0.050$&$0.048$&$0.047$&$0.048$&$0.048$&$0.046$&$0.046$&$0.049$&$0.049$\tabularnewline
 Han&$0.049$&$0.049$&$0.050$&$0.050$&$0.051$&$0.050$&$0.047$&$0.052$&$0.051$\tabularnewline
 Han ind&$0.047$&$0.048$&$0.050$&$0.047$&$0.048$&$0.049$&$0.046$&$0.046$&$0.048$\tabularnewline 
 \hline
\multicolumn{10}{c}{$\Pr(G=1) = 0.2$} \\ \hline
 U&$0.050$&$0.047$&$0.048$&$0.046$&$0.051$&$0.048$&$0.052$&$0.050$&$0.046$\tabularnewline
 U ind&$0.052$&$0.045$&$0.050$&$0.045$&$0.047$&$0.050$&$0.048$&$0.049$&$0.050$\tabularnewline
 prosp&$0.050$&$0.046$&$0.046$&$0.045$&$0.049$&$0.045$&$0.051$&$0.047$&$0.045$\tabularnewline
 Han&$0.051$&$0.048$&$0.050$&$0.048$&$0.050$&$0.048$&$0.055$&$0.051$&$0.050$\tabularnewline
 Han ind&$0.051$&$0.046$&$0.049$&$0.045$&$0.048$&$0.051$&$0.048$&$0.049$&$0.052$\tabularnewline
 \hline
\multicolumn{10}{c}{$\Pr(G=1) = 0.05$} \\ \hline
 U&$0.050$&$0.054$&$0.053$&$0.054$&$0.054$&$0.050$&$0.051$&$0.049$&$0.048$\tabularnewline
 U ind&$0.052$&$0.057$&$0.050$&$0.051$&$0.051$&$0.050$&$0.048$&$0.047$&$0.048$\tabularnewline
 prosp&$0.043$&$0.048$&$0.050$&$0.042$&$0.047$&$0.045$&$0.042$&$0.043$&$0.041$\tabularnewline
 Han&$0.052$&$0.048$&$0.050$&$0.050$&$0.054$&$0.052$&$0.051$&$0.051$&$0.050$\tabularnewline
 Han ind&$0.052$&$0.052$&$0.049$&$0.050$&$0.049$&$0.046$&$0.047$&$0.048$&$0.048$\tabularnewline
\hline
\end{tabular}
\caption{The type 1 error of the compared tests, under various combinations of the prevalence of the genetic variant $a_1$, and the effect of the genetic and environmental variables on the disease outcome ($\alpha_1$ and $\alpha_2$, respectively). The tests `U' and `U ind' are the proposed tests without and with the assumption of G-E independence. `prosp' is the usual test based on prospective likelihood, and `Han' and `Han ind' are the tests based on retrospective profile likelihood proposed by \cite{han2012likelihood}. The type 1 error was calculated from 10,000 simulations, each with 4000 cases and 4000 controls.  }
\end{table}

\section{Ovarian cancer application}

\noindent We applied the proposed test of additive interaction to the
well-known Israeli Ovarian Cancer data \cite{modan2001parity} also recently
analyzed by \cite{chatterjee2005semiparametric,tchetgen2010semiparametric,tchetgen22robust}. Although the goal in previous analyses
was to detect a multiplicative gene-environment interaction between having the
BRCA1/2 mutation and two environmental exposures, number of years of oral
contraceptive use (OC) and number of children (parity), here we are primarily
concerned with determining whether such interactions might be operating on the
additive scale. Both environmental exposures are naturally coded as counts,
and therefore can be modeled using Poisson regression, while standard logistic
regression was used to model the genetic variant. Both sets of models were
estimated only using controls as previously described. We present results when
assuming G-E independence, and without using such an assumption. Without G-E
independence, the odds ratio parameter $\omega$ was estimated as the
coefficient for the exposure in view in a logistic regression of the genetic
factor on the environmental exposure and covariates, i.e. $(E,X)$.

A number of covariates were available for confounding adjustment and also to
enforce the independence assumption. All regression models adjusted for age,
as an indicator variable for age$\leq50$, indicator variables for ethnicities
of Ashkenazi jew, and non-Ashkenazi (with mixed race serving as reference
category), indicator variables for personal history of breast cancer, family
history of breast cancer, and family history of ovarian cancer. For
convenience, we used the nonparametric bootstrap to evaluate 95\% confidence
intervals and p-values.

\begin{table}[pth]
\label{tab:res} \centering
\begin{tabular}
[c]{lccc}\hline
E variable & U & 95\% CI & p-values\\\hline
\multicolumn{4}{c}{G-E independence assumed}\\\hline
OC & 0.049 & (-0.006, 0.117) & 0.09\\
Parity & -0.044 & (-0.092, -0.004) & 0.03\\\hline
\multicolumn{4}{c}{No G-E independence assumed}\\\hline
OC & 0.002 & (-4.692, 0.012) & 0.77\\
Parity & -0.005 & (-0.023, 0.019) & 0.59\\\hline
\end{tabular}
\caption{Testing results for the additive G-E interaction between presence of
BRCA1/2 mutation (G) and number of years of oral contraceptive (OC) use, and
parity (E variables), with and without G-E independence assumption. U is the
proposed (standardized) test statistic, and its 95\% bootstrap confidence
interval and p-value are provided, calculated over 1000 bootstrap samples.}%
\end{table}

The table provides results from testing for a G$\times$E additive
interaction with and without making the G-E independence assumption. In
accordance with simulation results, the independence assumption yields a test
statistic consistently more extreme for both exposures in view than the
corresponding test which does not incorporate the assumption. Specifically, we
successfully reject the null hypothesis of no additive G-E interaction between
BRCA1/2 mutation and parity at the alpha level$=0.05$, only when the
independence assumption is made, and not otherwise. We found no conclusive
evidence of an additive interaction with OC, although the test statistic under
G-E independence was far more extreme than without the assumption and the
associated p-value was marginally significant (p-value=$0.09)$. \ It is
interesting to compare these findings with previous analyses of these data
that have primarily been concerned with detecting the presence of a
multiplicative G$\times$E interaction. For instance, \cite{tchetgen2010semiparametric} leveraged the independence assumption to detect a G$\times$E multiplicative
interaction only with OC and failed to find evidence of a similar interaction
with parity, thus essentially reporting the opposite findings to ours.
However, our findings are potentially more scientifically relevant given that
interactions on the multiplicative scale may be harder to interpret biologically.

\section{ Conclusion }

\noindent We have described a very general framework to test for G$\times$E additive
interactions exploiting G-E independence in case-control studies. The proposed
strategy has several advantages over existing RERI-based strategies, primarily
because, unlike the latter, the former does not require a regression model for
the outcome, and therefore is less vulnerable to model misspecification of the
outcome regression, a potential concern particularly if E is a count or
continuous and additional covariates are included in the regression. The
approach put forward in this paper is closely related to the semiparametric
framework of \cite{vansteelandt2008multiply}, who characterized the set of influence
functions of a model of interaction (on the additive or multiplicative scale)
under a semiparametric union model in which only a subset but not all of the
parametric models used to describe the data generating mechanism need to be
correct for valid inference. In fact, one can show that our proposed test
statistic belongs to the general class of test statistics for additive
interaction associated with their set of influence functions. However,
because \cite{vansteelandt2008multiply} did not allow for outcome dependent
sampling and only considered standard prospective random sampling, not all
test statistics in their class may be used under case-control sampling. Thus,
an important contribution of the current paper has been to characterize the
subset of the class of test statistics of an additive interaction that may be used both under prospective and retrospective sampling.

An important limitation of the proposed approach is that it does not readily
produce an estimate of the risk difference parameters which are often of
primary interest for understanding the public health significance of any
significant finding. To obtain such estimates, one would need an estimate of
the main effect of exposures, which are being treated as unspecified
nuisance parameters in the proposed approach from the ground of robustness.
Addressing this limitation is a priority for future research to extend the
methods describe herein.
\end{spacing}

\clearpage
\bibliography{GE}%

\begin{thebibliography}{10}

\bibitem{rothman1980concepts}
Rothman Kenneth~J, Greenland Sander, Walker Alexander~M. Concepts of
  interaction.  {\it American Journal of Epidemiology. }1980;112(4):467--470.

\bibitem{greenland1983tests}
Greenland Sander. Tests for interaction in epidemiologic studies: a review and
  a study of power.  {\it {Statistics in Medicine}. }1983;2(2):243--251.

\bibitem{cordell2002epistasis}
Cordell Heather~J. Epistasis: what it means, what it doesn't mean, and
  statistical methods to detect it in humans.  {\it Human Molecular Genetics.
  }2002;11(20):2463--2468.

\bibitem{vanderweele2014tutorial}
VanderWeele Tyler~J, Knol Mirjam~J. A tutorial on interaction.  {\it
  Epidemiologic Methods. }2014;3(1):33--72.

\bibitem{liu2017robust}
Liu Gang, Mukherjee Bhramar, Lee Seunggeun, et al. Robust tests for additive
  gene-environment interaction in case-control studies using gene-environment
  independence.  {\it American Journal of Epidemiology. }2017;187(2):366--377.

\bibitem{rothman2008modern}
Rothman Kenneth~J, others . {\it Modern epidemiology}.
\newblock Philadelphia, PA: Lippincott Williams \& Wilkins; 3~ed.2008.

\bibitem{piegorsch1994non}
Piegorsch Walter~W, Weinberg Clarice~R, Taylor Jack~A. Non-hierarchical
  logistic models and case-only designs for assessing susceptibility in
  population-based case-control studies.  {\it {Statistics in Medicine}.
  }1994;13(2):153--162.

\bibitem{umbach1997designing}
Umbach David~M, Weinberg Clarice~R. Designing and analysing case-control
  studies to exploit independence of genotype and exposure.  {\it {Statistics
  in Medicine}. }1997;16(15):1731--1743.

\bibitem{chatterjee2005semiparametric}
Chatterjee Nilanjan, Carroll Raymond~J. Semiparametric maximum likelihood
  estimation exploiting gene-environment independence in case-control studies.
  {\it Biometrika. }2005;92(2):399--418.

\bibitem{tchetgen2010semiparametric}
Tchetgen~Tchetgen Eric~J, Robins James. The semiparametric case-only estimator.
   {\it Biometrics. }2010;66(4):1138--1144.

\bibitem{tchetgen22robust}
Tchetgen~Tchetgen E. Robust Discovery of Genetic Associations incorporating
  Gene-Environment Interaction and Independence.(2011) Epidemiology.  {\it
  Volume. };22(2):262--272.

\bibitem{han2012likelihood}
Han Summer~S, Rosenberg Philip~S, Garcia-Closas Montse, et al. Likelihood Ratio
  Test for Detecting Gene (G)-Environment (E) Interactions Under an Additive
  Risk Model Exploiting G-E Independence for Case-Control Data.  {\it American
  Journal of Epidemiology. }2012;176(11):1060--1067.

\bibitem{skrondal2003interaction}
Skrondal Anders. Interaction as departure from additivity in case-control
  studies: a cautionary note.  {\it American Journal of Epidemiology.
  }2003;158(3):251--258.

\bibitem{knol2007estimating}
Knol Mirjam~J, Tweel Ingeborg, Grobbee Diederick~E, Numans Mattijs~E, Geerlings
  Mirjam~I. Estimating interaction on an additive scale between continuous
  determinants in a logistic regression model.  {\it International Journal of
  Epidemiology. }2007;36(5):1111--1118.

\bibitem{tchetgen2009doubly}
Tchetgen~Tchetgen Eric~J, Robins James~M, Rotnitzky Andrea. On doubly robust
  estimation in a semiparametric odds ratio model.  {\it Biometrika.
  }2010;97(1):171--180.

\bibitem{tchetgen2013general}
Tchetgen~Tchetgen Eric~J. A general regression framework for a secondary
  outcome in case--control studies.  {\it Biostatistics. }2013;15(1):117--128.

\bibitem{modan2001parity}
Modan Baruch, Hartge Patricia, Hirsh-Yechezkel Galit, et al. Parity, oral
  contraceptives, and the risk of ovarian cancer among carriers and noncarriers
  of a BRCA1 or BRCA2 mutation.  {\it New England Journal of Medicine.
  }2001;345(4):235--240.

\bibitem{vansteelandt2008multiply}
Vansteelandt Stijn, VanderWeele Tyler~J, Tchetgen Eric~J, Robins James~M.
  Multiply robust inference for statistical interactions.  {\it Journal of the
  American Statistical Association. }2008;103(484):1693--1704.

\end{thebibliography}
\clearpage
\appendix\noindent

\section*{Appendix}
\paragraph{Proof that }$Var(\sum_{i}\widehat{U}_{i}/n)>Var(\sum
_{i}\widetilde{U}_{i}/n).$

\noindent To show the result requires the influence function of
$\widehat{\theta}=\left(  \widehat{\omega},\widehat{p}_{1}=\widehat{p}%
_{1}(0),\widehat{p}_{2}=\widehat{p}_{2}(0)\right)  ^{T}$ which is of the form%

\[
IF=\mathbb{E}\left(  \frac{\partial R\left(  \theta\right)  }{\partial\theta
}\right)  ^{-1}R\left(  \theta\right)
\]
where $R\left(  \theta\right)  = \{(1-D)(1-A_2)\times[  (A_{1}-\mathbb{E}\left(
A_{1},\theta\right)  ],(1-D)(1-A_1)\times[A_{2}-\mathbb{E}\left(  A_{2};\theta\right)
],(1-D)\times[{{A_{1}A_{2}-\mathbb{E}\left(  A_{1}A_{2};\theta\right)}}  ]  \}^{T},$
where the first component is the score of $p_{1}(0),$ the second component is
the score of $p_{2}(0)$, the last component is the score of $\omega,$ and
$\theta=\left(  \omega,p_{1},p_{2}\right)  .$ Standard matrix algebra
can be used to show that at the submodel where $A_{1}$ and $A_{2}$ are
independent $IF=(IF_{1},IF_{2},IF_{3})$ where:%

\begin{align*}
IF_{1}  &  =\mathbb{E}\left[  \left(  1-A_{2}\right)  (1-D)\right]
^{-1}(1-A_{2})\left(  A_{1}-\mathbb{E}\left(  A_{1}\right)  \right)  (1-D)\\
&  \approx-\mathbb{E}\left[  \left(  1-\mathbb{E}\left(  A_{2}|D=0\right)
\right)  \right]  ^{-1}(A_{2}-\mathbb{E}\left(  A_{2}|D=0\right)  )\left(
A_{1}-\mathbb{E}\left(  A_{1}|D=0\right)  \right)  (1-D)\\
&  +\left(  A_{1}-\mathbb{E}\left(  A_{1}|D=0\right)  \right)  (1-D)\\
IF_{2}  &  =\mathbb{E}\left[  \left(  1-A_{1}\right)  (1-D)\right]
^{-1}(1-A_{1})\left(  A_{2}-\mathbb{E}\left(  A_{2}\right)  \right)
(1-D)\\
&  \approx-\mathbb{E}\left[  \left(  1-\mathbb{E}\left(  A_{1}|D=0\right)
\right)  \right]  ^{-1}(A_{1}-\mathbb{E}\left(  A_{1}|D=0\right)  )\left(
A_{2}-\mathbb{E}\left(  A_{2}|D=0\right)  \right)  (1-D)\\
&  +\left(  A_{2}-\mathbb{E}\left(  A_{2}|D=0\right)  \right)  (1-D)\\
IF_{3}  &  =\mathbb{E}\left[  \left(  A_{1}-\mathbb{E}\left(  A_{1}%
|D=0\right)  \right)  ^{2}|D=0\right]  ^{-1}\mathbb{E}\left[  \left(
A_{2}-\mathbb{E}\left(  A_{2}|D=0\right)  \right)  ^{2}|D=0\right]  ^{-1}\\
&  \times\left(  A_{1}-\mathbb{E}\left(  A_{1}|D=0\right)  \right)  \left(
A_{2}-\mathbb{E}\left(  A_{2}|D=0\right)  \right)  (1-D)
\end{align*}
A Taylor series argument then gives%
\begin{align*}
&  \sum_{i}\widehat{U}_{i}/\sqrt{n}\\
&  \approx\sum_{i}U_{i}/\sqrt{n}-\mathbb{E}\left[  \left(  A_{2}%
-p_{2}(0)\right)  D\right]  IF_{1}\\
&  -\mathbb{E}\left[  \left(  A_{1}-p_{1}(0)\right)  D\right]  IF_{2}%
-\mathbb{E}\left[  A_{1}A_{2}\left(  A_{2}-p_{2}(0)\right)  \left(
A_{1}-p_{1}(0)\right)  D\right]  IF_{3}\\
&  =\sum_{i}U_{i}/\sqrt{n}\\
&  -\mathbb{E}\left[  \left(  A_{2}-p_{2}(0)\right)  D\right]  \sum_{i}\left(
A_{1,i}-\mathbb{E}\left(  A_{1}|D=0\right)  \right)  (1-D_{i})/\sqrt{n}\\
&  -\mathbb{E}\left[  \left(  A_{1}-p_{1}(0)\right)  D\right]  \sum_{i}\left(
A_{2,i}-\mathbb{E}\left(  A_{2}|D=0\right)  \right)  (1-D_{i})/\sqrt{n}\\
&  -\left(
\begin{array}
[c]{c}%
\mathbb{E}\left[  A_{1}A_{2}\left(  A_{2}-p_{2}\right)  \left(  A_{1}%
-p_{1}\right)  D\right]  \left\{  p_{1}p_{2}\left(  1-p_{1}\right)  \left(
1-p_{2}\right)  \right\}  ^{-1}\\
+\mathbb{E}\left[  \left(  A_{2}-p_{2}\right)  D\right]  \left[  \left(
1-p_{2}\right)  \right]  ^{-1}+\mathbb{E}\left[  \left(  A_{1}-p_{1}\right)
D\right]  \left[  \left(  1-p_{1}\right)  \right]  ^{-1}%
\end{array}
\right) \\
&  \times\sum_{i}\left(  A_{1,i}-\mathbb{E}\left(  A_{1}|D=0\right)  \right)
\left(  A_{2,i}-\mathbb{E}\left(  A_{2}|D=0\right)  \right)  (1-D_{i}%
)/\sqrt{n}%
\end{align*}

Upon noting that the above four terms are mutually uncorrelated, we have that
:%
\[
Var\left(  \sum_{i}\widehat{U}_{i}/n\right)  \approx V_{1}+V_{2}+V_{3}%
\]
where
\begin{align*}
V_{1}  &  =Var\left(  U\right)  /n\\
V_{2}  &  =\mathbb{E}\left[  \left(  A_{2}-p_{2}(0)\right)  D\right]
^{2}Var\left(  \left(  A_{1}-\mathbb{E}\left(  A_{1}|D=0\right)  \right)
(1-D)\right)  /n\\
&  +\mathbb{E}\left[  \left(  A_{1}-p_{1}(0)\right)  D\right]  ^{2}Var\left(
\left(  A_{2}-\mathbb{E}\left(  A_{2}|D=0\right)  \right)  (1-D)\right)  /n\\
V_{3}  &  =\left(
\begin{array}
[c]{c}%
\mathbb{E}\left[  A_{1}A_{2}\left(  A_{2}-p_{2}\right)  \left(  A_{1}%
-p_{1}\right)  D\right]  \left\{  p_{1}p_{2}\left(  1-p_{1}\right)  \left(
1-p_{2}\right)  \right\}  ^{-1}\\
+\mathbb{E}\left[  \left(  A_{2}-p_{2}\right)  D\right]  \left[  \left(
1-p_{2}\right)  \right]  ^{-1}+\mathbb{E}\left[  \left(  A_{1}-p_{1}\right)
D\right]  \left[  \left(  1-p_{1}\right)  \right]  ^{-1}%
\end{array}
\right)  ^{2}\\
&  \times Var\left(  \left(  A_{1}-\mathbb{E}\left(  A_{1}|D=0\right)
\right)  \left(  A_{2}-\mathbb{E}\left(  A_{2}|D=0\right)  \right)
(1-D)\right)  /n
\end{align*}
A similar derivation shows that
\begin{align*}
&  \sum_{i}\widetilde{U}_{i}/\sqrt{n}\\
&  \approx\sum_{i}U_{i}/\sqrt{n}\\
&  -\mathbb{E}\left[  \left(  A_{2}-p_{2}(0)\right)  D\right]  \sum_{i}\left(
A_{1,i}-\mathbb{E}\left(  A_{1}|D=0\right)  \right)  (1-D_{i})/\sqrt{n}\\
&  -\mathbb{E}\left[  \left(  A_{1}-p_{1}(0)\right)  D\right]  \sum_{i}\left(
A_{2,i}-\mathbb{E}\left(  A_{2}|D=0\right)  \right)  (1-D_{i})/\sqrt{n}%
\end{align*}
which gives
\[
Var\left(  \sum_{i}\widetilde{U}_{i}/n\right)  \approx V_{1}+V_{2}%
\]
proving the result.
\\
\paragraph{Asymptotic variance for unified class of test statistics }

\noindent Our proposed test statistic is then given by $Z=\sum_{i}%
\widehat{W}_{i}(g)/n\widehat{\sigma}_{W},$ where $\widehat{\sigma}_{W}^{2}$ is
an estimate of $Var\left(  \sum_{i}\widehat{W}_{i}(g)/n\right)  $ one can
derive using a standard Taylor series argument:
\begin{equation}
Var\left(  \sum_{i}\widehat{W}_{i}(g)/n\right)  \approx n^{-1}Var\left(
W(g,\theta)\right)  +n^{-1}\mathbb{E}\left(  W_{\theta}^{T}(g)\right)
Var\left(  S_{\theta}^{\dag}\right)  \mathbb{E}\left(  W_{\theta}(g)\right)
\label{Asy. Variance}%
\end{equation}
where $W_{\theta}(g)$ is the derivative of $W(g,\theta)$ with respect to
$\theta$ evaluated at the truth, and $S_{\theta}^{\dag}$ is the influence
function of $\widehat{\theta}$ \cite{tchetgen2009doubly}. For instance, when $\widehat{\theta}$ is a
maximum likelihood estimator, $S_{\theta}^{\dag}=\mathbb{E}\left(  S_{\theta
}S_{\theta}^{T}\right)  ^{-1}S_{\theta},$ where $S_{\theta}$ denote the score
of $\theta.$ Under the assumption that $A_{1}$ and $A_{2}$ are independent, we
may set $\widehat{\omega}=1$ and redefine $\theta=\left(  \alpha_{1}%
,\alpha_{2}\right)  ,$also note that under independence, the joint density
$\left(  3\right)  $ simplifies to $f(A_{1},A_{2}|X)=f(A_{1}|X)f(A_{2}|X),$
leading to some simplification in the above expression for the asymptotic
variance of the test statistic.

\paragraph{Proof of Lemma 1. }\textit{Consider the nonparametric
	additive representation of }$\mu(a_{1},a_{2},x)$ \textit{given by }$\mu
(a_{1},a_{2},x)=\beta_{1}(a_{1},x)+\beta_{2}(a_{2},x)+\beta_{3}(a_{1}%
,a_{2},x)+\beta_{4}(x)$\textit{ where }$\beta_{1}(a_{1},x)$\textit{ is the
	main effect of }$A_{1}$\textit{ and satisfies }$\beta_{1}(0,x)=0,$\textit{
	likewise }$\beta_{2}(a_{2},x)$\textit{ is the main effect of }$A_{2}$\textit{
	and satisfies }$\beta_{2}(0,x)=0,\beta_{3}(a_{1},a_{2},x)$\textit{ is the
	additive interaction between }$A_{1}$\textit{ and }$A_{2}$\textit{ and
	satisfies }$\beta_{3}(0,a_{2},x)=\beta_{3}(a_{1},0,x)=0,$\textit{ and }%
$\beta_{4}(x)$\textit{ is the main effect of }$X.$\textit{ For any function
}$g,$\textit{ note that }%
\begin{align*}
&  \mathbb{E}\left\{  W(g)|D=1,x\right\}  \\
&  =\int\int w(a_{1},a_{2},x,1;g)\mu(a_{1},a_{2},x)f(a_{1},a_{2}%
|x)f(x)d\nu(a_{1},a_{2})/\int\int\mu(a_{1},a_{2},x)f(a_{1},a_{2}%
|x)f(x)d\nu(a_{1},a_{2})\\
&  \propto\int\int w(a_{1},a_{2},x,1;g)\mu(a_{1},a_{2},x)f(a_{1}%
|A_{2}=0,x)f(a_{2}|A_{1}=0,x)OR(a_{1},a_{2};x)d\nu(a_{1},a_{2})\\
&  =\int\int w(a_{1},a_{2},x,1;g)\beta_{3}(a_{1},a_{2},x)f(a_{1}%
|A_{2}=0,x)f(a_{2}|A_{1}=0,x)OR(a_{1},a_{2};x)d\nu(a_{1},a_{2})
\end{align*}
\textit{since
	\begin{align*}
	& \int w\left(  a_{1},a_{2},x,1;g\right)  \beta_{1}\left(  a_{1},x\right)
	f\left(  a_{2}|A_{1}=0,x\right)  OR\left(  a_{1},a_{2};x\right)  d\nu\left(
	a_{2}\right)  \\
	& =\{\beta_{1}(a_{1},x)\int g(A_{1},A_{2},X)f(a_{2}|A_{1}=0,x)d\nu(a_{2})\\
	& -\beta_{1}\left(  a_{1},x\right)  \int g\left(  A_{1},a_{2},X\right)
	f\left(  a_{2}|A_{1}=0,X\right)  d\mu(a_{2})\\
	& -\beta_{1}(a_{1},x)\int\int g\left(  a_{1},A_{2},X\right)  f\left(
	a_{1}|A_{2}=0,X\right)  d\mu(a_{1})f\left(  a_{2}|A_{1}=0,X\right)d\nu(a_{2})\\
	& +\beta_{1}(a_{1},x)\int\int g\left(  a_{1},a_{2},X\right)  f\left(
	a_{1}|A_{2}=0,X\right)f\left(  a_{2}|A_{1}=0,X\right)  d\mu(a_{1})d\mu(a_{2})\\
	& =0
	\end{align*}
	furthermore by symmetry,}%
\begin{align*}
& \int w\left(  a_{1},a_{2},x,1;g\right)  \beta_{2}\left(  a_{2},x\right)
f\left(  a_{1}|A_{2}=0,x\right)  OR\left(  a_{1},a_{2};x\right)  d\nu\left(
a_{1}\right)  \\
& =0
\end{align*}
\textit{and finally}  \textit{%
	\begin{align*}
	& \int w\left(  a_{1},a_{2},x,1;g\right)  \beta_{4}(x)f\left(  a_{2}%
	|A_{1}=0,x\right)  OR\left(  a_{1},a_{2};x\right)  d\nu\left(  a_{2}\right)
	\\
	& =\{\beta_{4}(x)\int g(A_{1},A_{2},X)f(a_{2}|A_{1}=0,x)d\nu(a_{2})\\
	& -\beta_{4}(x)\int g\left(  A_{1},a_{2},X\right)  f\left(  a_{2}%
	|A_{1}=0,X\right)  d\mu(a_{2})\\
	& -\beta_{4}(x)\int\int g\left(  a_{1},A_{2},X\right)  f\left(  a_{1}%
	|A_{2}=0,X\right)  d\mu(a_{1})f\left(  a_{2}|A_{1}=0,X\right)d\nu(a_{2})\\
	& +\beta_{4}(x)\int\int g\left(  a_{1},A_{2},X\right)  f\left(  a_{1}%
	|A_{2}=0,X\right)f\left(  a_{2}|A_{1}=0,X\right)  d\mu(a_{1})d\mu(a_{2})\\
	& =0
	\end{align*}
	therefore}%
\begin{align*}
& \int\int w(a_{1},a_{2},x,1;g)\left\{  \beta_{1}(a_{1},x)+\beta_{2}%
(a_{2},x)+\beta_{4}(x)\right\}  \\
& \times f\left(  a_{1}|A_{2}=0,x\right)  f\left(  a_{2}|A_{1}=0,x\right)
d\nu(a_{1},a_{2})\\
& =0
\end{align*}
\textit{for any choice of }$g.$\textit{ Thus, the null of no additive
	interaction }$\beta_{3}(a_{1},a_{2},x)=0$\textit{ for all }$\left(
a_{1},a_{2},x\right)  $\textit{ implies that }$E\left\{  W(g)|D=1,x\right\}
=0.$\textit{ We get the result in the other direction by choosing }%
$g(a_{1},a_{2},x)=g^{\ast}(a_{1},a_{2},x)=\beta_{3}(a_{1},a_{2},x)$\textit{
	which gives }%
\begin{align*}
\mathbb{E}\left\{  W(g)|D=1,x\right\}   &  =0\text{ \textit{for all}
}\mathit{g}\text{ \textit{and }}\mathit{x}\text{ \textit{implies that } }\\
\int\int w(a_{1},a_{2},x,1;g^{\ast})^{2}f(a_{1}|A_{2} &  =0,x)f(a_{2}%
|A_{1}=0,x)d\nu(a_{1},a_{2})=0\text{ \textit{for all} }x
\end{align*}
\textit{which in turn implies that }$\beta_{3}(a_{1},a_{2},x)=0$\textit{ for
	all }$\left(  a_{1},a_{2},x\right)  $\textit{ proving the result. }$\square$

\clearpage
\begin{figure}[htp]
	\begin{centering}\label{fig}
		\makebox[\textwidth][c]{\includegraphics[width=1\textwidth, scale=.4]{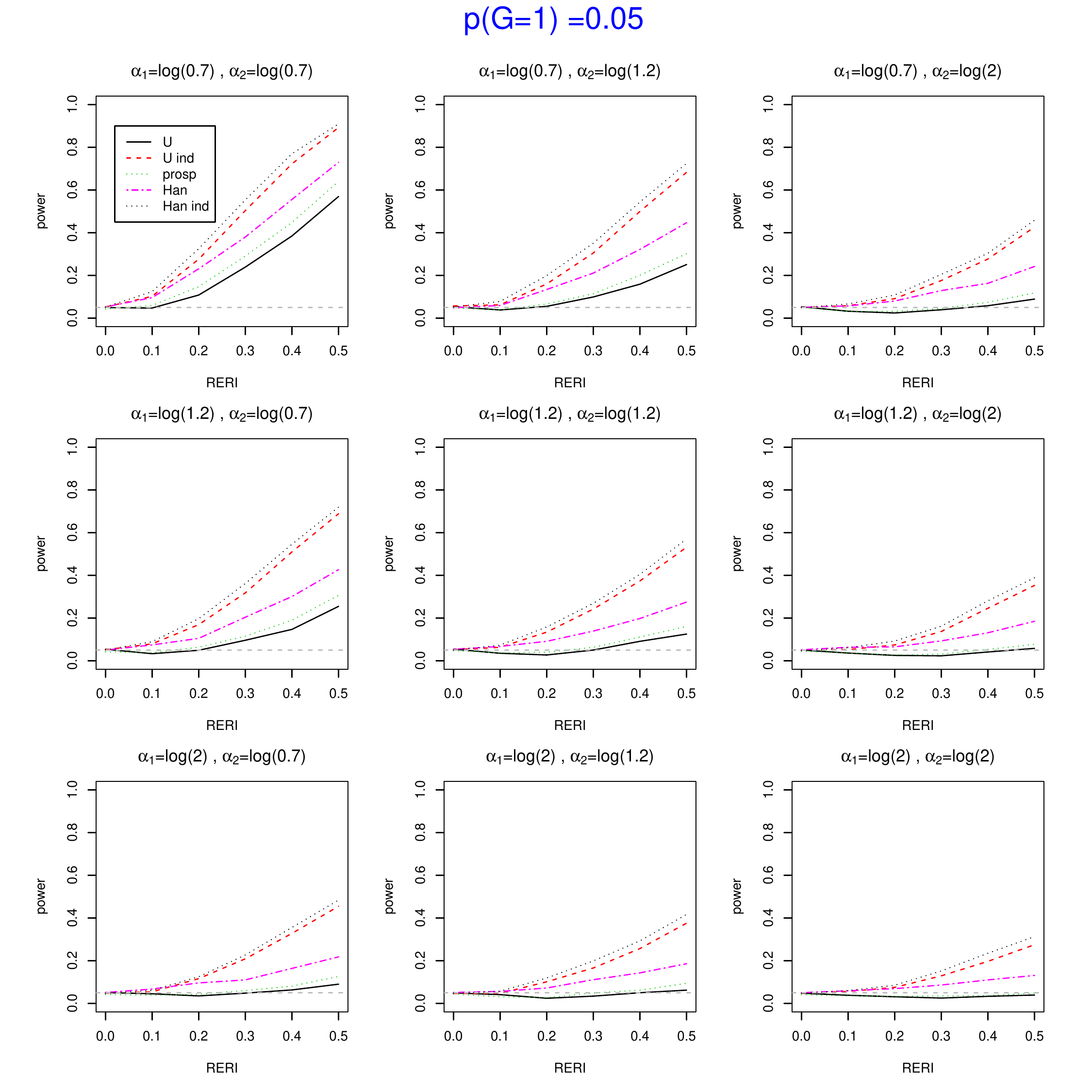}}
	\end{centering}
	\caption{Power of the various tests for identifying additive G-E interaction, in the simple settings in which both $a_1$ and $a_2$ are binary, and no covariates are used in the model. The simulated $a_1$ is common $\Pr(a_1=1) =0.05$, $\Pr(a_2=1) =0.2$, and the disease model $\Pr(D=1|a_1,a_2)$ has the fixed baseline risk $\alpha_0 =\mbox{logit}(0.01)$, the main effects of the $a_1, a_2$ are varied as $\alpha_1,\alpha_2$, and the RERI is varied from 0 to 0.5. The compared estimators are the proposed `U' and `U ind' (without and with assuming G-E independence), `prosp' is the standard test under prospective logistic regression and `Han' and `Han ind' is the retrospective profile likelihood test proposed in Han (2012) without and with assuming G-E independence.
	}
\end{figure}
\begin{figure}[htp]
	\begin{centering}\label{fig}
		\makebox[\textwidth][c]{\includegraphics[width=1\textwidth, scale=.4]{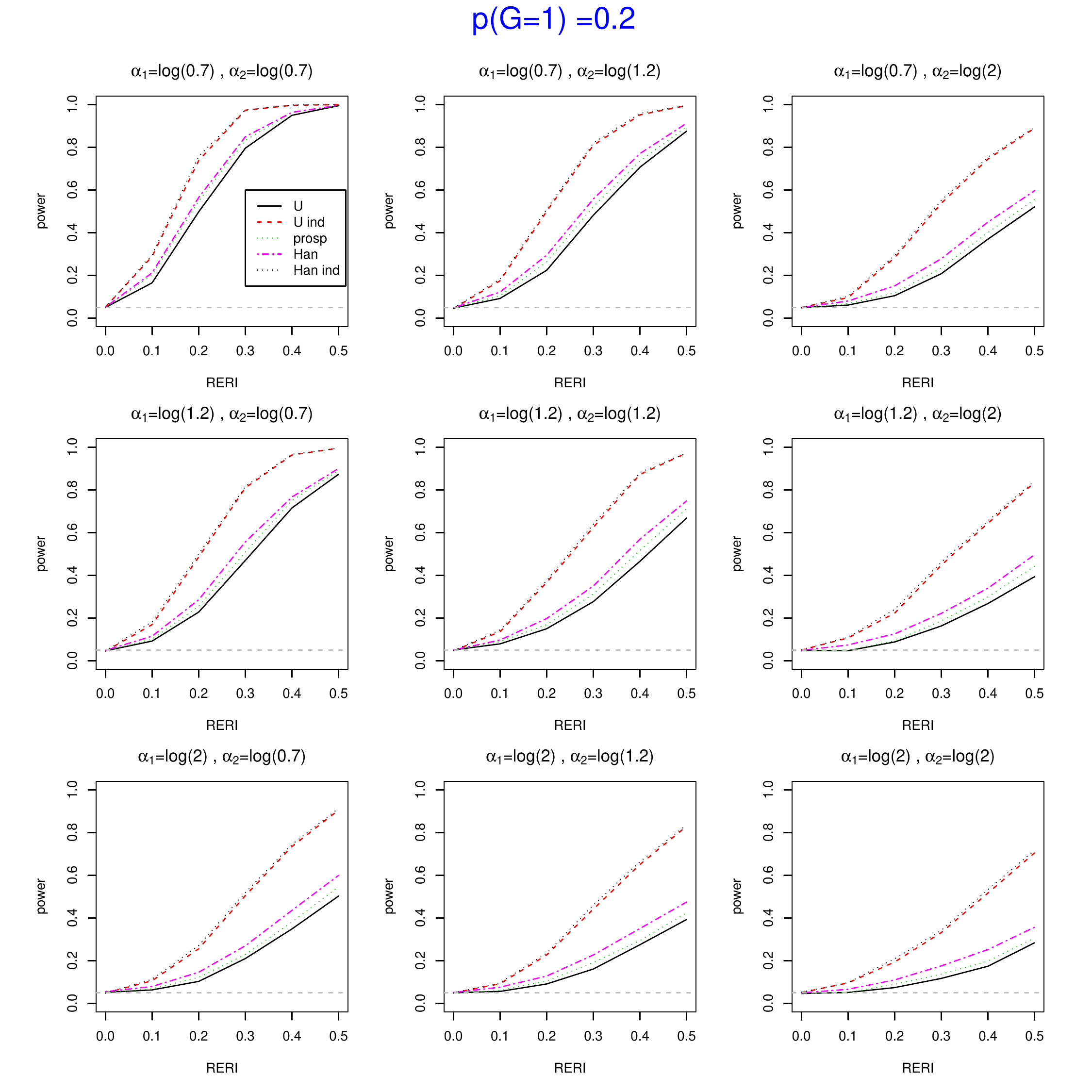}}
	\end{centering}
	\caption{Power of the various tests for identifying additive G-E interaction, in the simple settings in which both $a_1$ and $a_2$ are binary, and no covariates are used in the model. The simulated $a_1$ is common $\Pr(a_1=1) =0.2$, $\Pr(a_2=1) =0.2$, and the disease model $\Pr(D=1|a_1,a_2)$ has the fixed baseline risk $\alpha_0 =\mbox{logit}(0.01)$, the main effects of the $a_1, a_2$ are varied as $\alpha_1,\alpha_2$, and the RERI is varied from 0 to 0.5. The compared estimators are the proposed `U' and `U ind' (without and with assuming G-E independence), `prosp' is the standard test under prospective logistic regression and `Han' and `Han ind' is the retrospective profile likelihood test proposed in Han (2012) without and with assuming G-E independence.
	}
\end{figure}

\end{document}